\documentclass{jpp}
\usepackage{graphicx}
\graphicspath{{./}{figures/}}

\usepackage[utf8]{inputenc}
\usepackage[T1]{fontenc}
\usepackage{amsmath}
\usepackage{bm}
\usepackage{xcolor}

\newcommand{\grad}[1]{\nabla #1}
\newcommand{\divg}[1]{\nabla \cdot #1}

\shorttitle{Compressible Tearing Mode}
\shortauthor{C. Shi and M. Velli}

\title{Linear Growth Rate of Compressible Tearing Instability in a Force-Free Current Sheet}

\author{Chen Shi\aff{1}
  \corresp{\email{chenshi@auburn.edu}},
  \and Marco Velli\aff{2}}

\affiliation{\aff{1}Department of Physics, Auburn University, Auburn, Alabama 36849, USA
\aff{2}Department of Earth, Planetary, and Space Sciences, University of California, Los Angeles, Los Angeles, CA 90095, USA}

\begin{document}

\maketitle

\begin{abstract}
We investigate the linear growth rate of tearing instability in a resistive force-free current sheet using the linearized fully-compressible magnetohydrodynamics (MHD) equations.
Our results show that, as the plasma beta ($\beta$), which is proportional to the ratio of thermal pressure to magnetic pressure, increases, the dispersion relation approaches that of the incompressible tearing mode. 
As $\beta$ decreases, the growth rate increases but this increase remains finite even in the limit $\beta \rightarrow 0$.
The relative increase in the growth rate from $\beta = +\infty$ to $\beta=0$ depends on the Lundquist number, with lower Lundquist numbers showing a more pronounced increase.
These findings indicate that, while compressibility has a negligible effect on the growth rate of tearing instability in high Lundquist number regimes, its effect becomes significant at low Lundquist numbers.
This suggests that compressibility may influence the critical Lundquist number for the transition from laminar to plasmoid-dominated magnetic reconnection.
Moreover, in low-$\beta$ regimes ($\beta < 1$), the generated density perturbation can be substantial, comparable to or even larger than the magnetic field perturbation, even though the growth rate is only slightly modified by the compressibility.
\end{abstract}

\section{Introduction}\label{sec:intro}
The tearing mode is a fundamental instability of resistive plasma current sheets. 
It provides a major mechanism for the rapid disruption and dissipation of current sheets on magnetohydrodynamic (MHD) scales.
Understanding the tearing instability is therefore crucial for explaining instabilities in Tokamak plasmas \citep{rutherford1973nonlinear,white1977saturation,connor1988tearing}, as well as explosive phenomena in space and astrophysical plasmas, including coronal mass ejections \citep{lin2015review}, solar flares \citep{lu2022observational}, and nanoflares \citep{jafari2021nanoflare}.

The tearing mode was first analyzed by \citet{furth1963finite} and \citet{coppi1976resistive}.
In an infinitely long current sheet, the growth rate of the most unstable tearing mode, $\gamma_m$, scales as $\gamma_m \tau \sim S^{-1/2}$, where $\tau = a/V_A$ is the Alfv\'en crossing time, $a$ is the current sheet thickness, and $V_A $ is the upstream Alfv\'en speed. The Lundquist number is defined as $S = aV_A/\eta$, where $\eta$ is the magnetic diffusivity.
In most space and laboratory plasma environments, the particle collision frequency is low and $\eta$ is consequently very small. The corresponding Lundquist number is therefore typically very large \citep{ji2011phase}, implying an extremely small tearing-mode growth rate.

Over the past two decades, it has been recognized that the tearing growth rate in a macroscopic current sheet depends not only on the Lundquist number but also on the current-sheet aspect ratio. 
For a two-dimensional current sheet of finite length $L$, the maximum growth rate expressed in terms of the macroscopic Alfv\'en time is $\gamma_m \tau_L \sim S_L^{-1/2} \left(a/L\right)^{-3/2}$, where $S_L= LV_A/\eta$ and $\tau_L = L/V_A$. 
Thus, the tearing growth rate is controlled jointly by the Lundquist number and the current sheet aspect ratio \citep{shibata2001plasmoid,loureiro2007instability,uzdensky2010fast}.

In particular, as a current sheet thins and its aspect ratio approaches $a/L \sim S_L^{-1/3}$, the tearing mode growth rate becomes independent of $S_L$ and is of order $\tau_L^{-1}$. This regime is known as ``ideal'' tearing \citep{pucci2013reconnection,del2016ideal,papini2018fast}.
Numerical simulations have confirmed that the tearing instability can rapidly disrupt a long, thin current sheet into a large number of plasmoids \citep{bhattacharjee2009fast,huang2013plasmoid,tenerani2015magnetic,shi2019fast}, enhancing the reconnection rate.

More recently, tearing-mode theory has been extended to incorporate a variety of physical effects, including Hall physics \citep{pucci2017fast,papini2019fast}, guide fields \citep{shi2020oblique}, different equilibrium magnetic field configurations \citep{pucci2018onset,shi2021stability}, neutral particles \citep{pucci2020tearing}, and plasma jets \citep{shi2022instabilities}. 
Many of these theoretical models adopt the incompressible approximation. 
Nevertheless, the influence of compressibility on the tearing instability has also been considered in several earlier studies.
Numerical calculations by \citet{tachi1985radiative} examined the compressible tearing mode using a linearized radiative-MHD code and found that compressibility slightly reduces the tearing growth rate relative to the incompressible case.
Recent work by \citet{de2024modification} analyzed compressible linear tearing modes using the LEGOLAS eigenvalue solver \citep{de2022legolas} and found that the tearing mode growth rate increases with plasma-$\beta$, implying that incompressible tearing mode has larger growth rate than the compressible tearing mode. 
We note, however, that \citet{tachi1985radiative} considered only the $\beta=0.01$ case, whereas \citet{de2024modification} examined the $\beta$-dependence at only a single wavenumber.

Two-dimensional compressible-MHD simulations have also shown that compressibility can modify the critical Lundquist number for the onset of plasmoid instability \citep{ni2012effects,baty2014effect}.
In particular, these simulations generally find a lower critical Lundquist number at higher plasma-$\beta$, suggesting enhanced tearing instability at higher $\beta$, in qualitative agreement with the trend reported by \citet{de2024modification}.

Despite these previous efforts, a systematic investigation of the plasma-$\beta$ dependence of the linear compressible tearing mode over a wide wavenumber range is still lacking.
An independent and systematic quantitative analysis of the linear eigenvalue problem therefore remains valuable.
In this study, we derive the linearized compressible-MHD equation set for a force-free current sheet, and calculate the linear growth rate of tearing mode by solving the resulting boundary-value-problem. 
This approach provides an independent determination of the linear eigenvalues and eigenfunctions and enables us to quantify systematically the effects of compressibility under the assumptions considered here. We further compare the compressible results with their incompressible counterparts and characterize the dependence of the growth rate and perturbation structure on plasma-$\beta$.
The remainder of the paper is organized as follows. In Section~\ref{sec:equations}, we derive the governing equation set and boundary conditions. In Section~\ref{sec:results} we present the numerical results. In Section~\ref{sec:conclusion}, we summarize this study and discuss the implications of our results.

\section{Derivation of equations to solve}\label{sec:equations}

\begin{figure}
    \centering
    \includegraphics[width=0.3\linewidth]{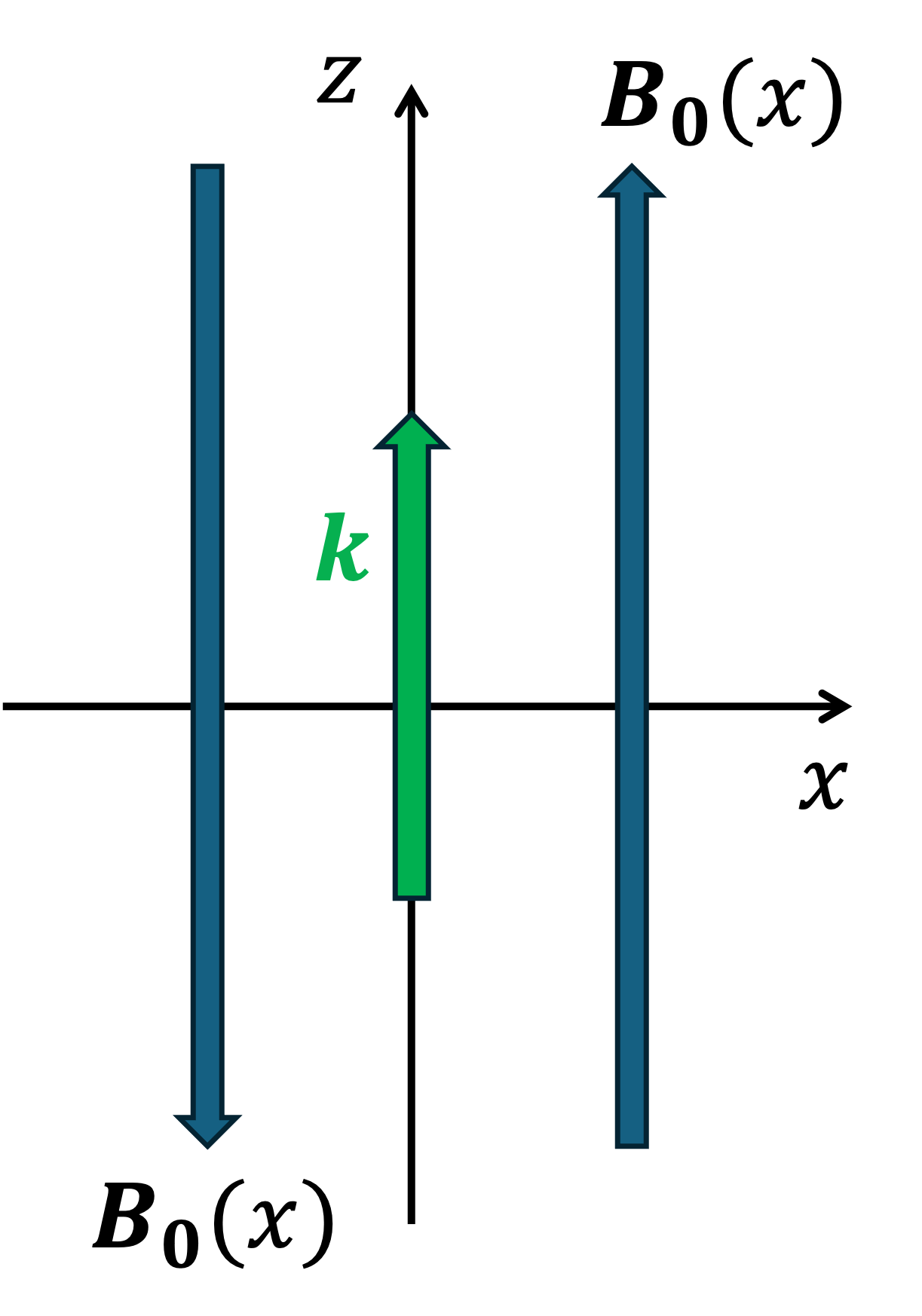}
    \caption{Coordinate system adopted in this study.}
    \label{fig:coordinates}
\end{figure}

\subsection{Background fields}
We start from the compressible resistive-MHD equation set with an adiabatic closure:
\begin{subequations}\label{eq:full_compressible_MHD}
    \begin{equation}
        \frac{\partial \rho}{\partial t} + \nabla \cdot \left( \rho \bm{U} \right) = 0
    \end{equation}
    \begin{equation}
        \rho \left( \frac{\partial \bm{U}}{\partial t} + \bm{U} \cdot \nabla \bm{U} \right) = - \nabla \left( P + \frac{B^2}{2 \mu_0} \right) + \frac{1}{\mu_0} \bm{B} \cdot  \nabla \bm{B} 
    \end{equation}
    \begin{equation}
        \frac{\partial \bm{B}}{\partial t} = \nabla \times \left( \bm{U}  \times \bm{B} \right) + \eta \nabla^2 \bm{B}
    \end{equation}
    \begin{equation}
       \frac{\partial P}{\partial t} = -\bm{U} \cdot \nabla P - \kappa  P \nabla \cdot \bm{U} 
    \end{equation}
\end{subequations}
where $\rho$, $\bm{U}$, $\bm{B}$, and $P$ denote density, velocity, magnetic field, and thermal pressure respectively; $\mu_0$ is the vacuum magnetic permeability, and $\kappa$ is the adiabatic index.

We note that, in the solar corona and lower solar atmosphere, various thermodynamic processes, such as thermal conduction and radiative cooling, may become important and thus invalidate the adiabatic closure adopted here. A detailed investigation of these effects, however, is beyond the scope of the present study.

As illustrated in Fig.~\ref{fig:coordinates}, we assume that $z$-axis is aligned with the wave vector ($\bm{k}$) direction and that the background fields vary only along the $x$-direction.
Here, we consider a force-free current sheet in which the background density $\rho_0$ and pressure $P_0$ are uniform, the background velocity vanishes everywhere, and the background magnetic field takes the form
\begin{equation}
    \bm{B_0}(x) = B_0 \, \tanh{\left( x/a \right)} \hat{e}_z + B_0 \, \mathrm{sech} \left( x/a \right) \hat{e}_y,
\end{equation}
so that $\left| \bm{B_0} \right| = Const$ everywhere. 
We note that the background field is in exact equilibrium only when $\eta=0$.
For weakly-collisional plasmas, $\eta \rightarrow 0$ and the tearing mode growth timescale is much shorter than the diffusion timescale of background field. We can therefore treat the background field as stationary.
Strictly, the stationary-background approximation requires the tearing growth time to be shorter than the resistive diffusion time of the equilibrium. 
This condition becomes marginal or invalid at the smallest Lundquist numbers considered here. 
Results in this regime should therefore be regarded as properties of the corresponding stationary eigenvalue problem rather than of a self-consistently evolving resistive current sheet.

\subsection{Equations for perturbations}
We denote the perturbations in density, velocity, and magnetic field by $\rho_1$, $\bm{u}$, $\bm{b}$ respectively.
The pressure perturbation is simply
\begin{displaymath}
    p_1 = C_s^2 \rho_1
\end{displaymath}
where $C_S = \sqrt{\kappa P_0 /\rho_0}$ is the sound speed.
Linearization and normalization Equation~(\ref{eq:full_compressible_MHD}) gives
\begin{subequations}\label{eq:linearized_general}
    \begin{equation}
        \gamma \rho = - \divg{\bm{u}} 
    \end{equation}
    \begin{equation}
        \gamma \bm{u} = - \beta \grad{\rho}   -\nabla \left( \bm{B} \cdot \bm{b}\right) + \bm{B}\cdot \nabla \bm{b} + \bm{b} \cdot \nabla \bm{B} 
    \end{equation}
    \begin{equation}
        \gamma \bm{b} = -\bm{u} \cdot \nabla \bm{B} + \bm{B} \cdot \nabla \bm{u} - \bm{B} \left( \nabla \cdot \bm{u} \right) + \frac{1}{S} \nabla^2 \bm{b}
    \end{equation}
\end{subequations}
where $\rho = \rho_1/\rho_0$. The magnetic field is expressed in Alfv\'en-speed unit through $B \rightarrow B/\sqrt{\mu_0 \rho_0}$, all velocities are normalized by the upstream Alfv\'en speed $V_A = B_0 / \sqrt{\mu_0 \rho_0}$, and all lengths are normalized by the current sheet thickness $a$.
$\beta = C_s^2 / V_A^2$ is the plasma beta, which is $\kappa/2$ times the conventional definition $P_0/(B_0^2/2\mu_0)$. $S = a V_A / \eta $ is the Lundquist number.

For the infinitely long current sheet in the $z$ direction considered here, the perturbations take the form
\begin{equation}
    \left( \begin{array}{c}
         \rho(x,z,t)  \\
         \bm{u}(x,z,t) \\
         \bm{b}(x,z,t)
    \end{array} \right)  =  \left( \begin{array}{c}
         \rho(x)  \\
         \bm{u}(x) \\
         \bm{b}(x)
    \end{array} \right)  e^{\gamma t + i kz}.
\end{equation}
Substituting these expressions into Equation~(\ref{eq:linearized_general}), we obtain the following ordinary-differential-equation set
\begin{subequations}\label{eq:full_ODE}
\begin{equation}\label{eq:full_ODE_bx}
    \gamma b_x = - k B_z u_x + \frac{1}{S} \left( b_x^{\prime \prime} - k^2 b_x \right)
\end{equation}
\begin{equation}\label{eq:full_ODE_by}
    \gamma b_y = - B_y^\prime u_x + kB_z u_y - B_y \left(u_x^\prime + k u_z \right) + \frac{1}{S} \left( b_y^{\prime\prime} - k^2 b_y \right) 
\end{equation}
\begin{equation}\label{eq:full_ODE_ux}
    \gamma u_x = \frac{\beta}{\gamma}\left( u_x^{\prime\prime} + k u_z^\prime \right) 
    - \left( B_z \frac{b_x^\prime}{k}  + B_y b_y \right)^\prime +  k B_z b_x
\end{equation}
\begin{equation}\label{eq:full_ODE_uy}
    \gamma u_y =  - k B_z b_y + B_y^\prime b_x 
\end{equation}
\begin{equation}\label{eq:full_ODE_uz}
   \gamma u_z = - \frac{k\beta}{\gamma} \left( u_x^\prime + k u_z \right) +k B_y b_y + B_z^\prime b_x
\end{equation}
\end{subequations}
where the prime denotes $d/dx$. We have eliminated $\rho$ using the continuity equation and eliminated $b_z$ using the divergence-free condition $\divg{\bm{b}} = 0$.
We also adopt the mappings: $i b_x \rightarrow b_x$, $i u_y \rightarrow u_y$, and $i u_z \rightarrow u_z$ so that the equation set becomes purely real. Accordingly, we restrict our analysis to the purely growing tearing branch and therefore take $\gamma$ to be real, as in the incompressible case.
We note that Equation~(\ref{eq:full_ODE_uz}) can be used to eliminate the term $\frac{\beta}{\gamma}\left( u_x^{\prime\prime} + k u_z^\prime \right)$ in Equation~(\ref{eq:full_ODE_ux}).

In the limit $\beta \rightarrow +\infty$, Equation~(\ref{eq:full_ODE_uz}) reduces to $u_x^\prime + k u_z = 0$, i.e. the incompressible condition $\divg{\bm{u}} = 0$.
In this case, Equation~(\ref{eq:full_ODE_ux}) can be simplified to 
\begin{displaymath}
    \frac{\gamma}{k} \left( u_x^{\prime \prime} - k^2 u_x \right) = B_z \left( b_x^{\prime\prime} - k^2 b_x \right) - B_z^{\prime\prime} b_x,
\end{displaymath}
which, together with Equation~(\ref{eq:full_ODE_bx}), forms the two-equation set for the classical incompressible tearing instability \cite{pucci2013reconnection}.
In the limit $\beta \rightarrow0$, the order of the equation set is reduced because $u_x^{\prime\prime}$ terms vanish.  

\begin{figure}
    \centering
    \includegraphics[width=\linewidth]{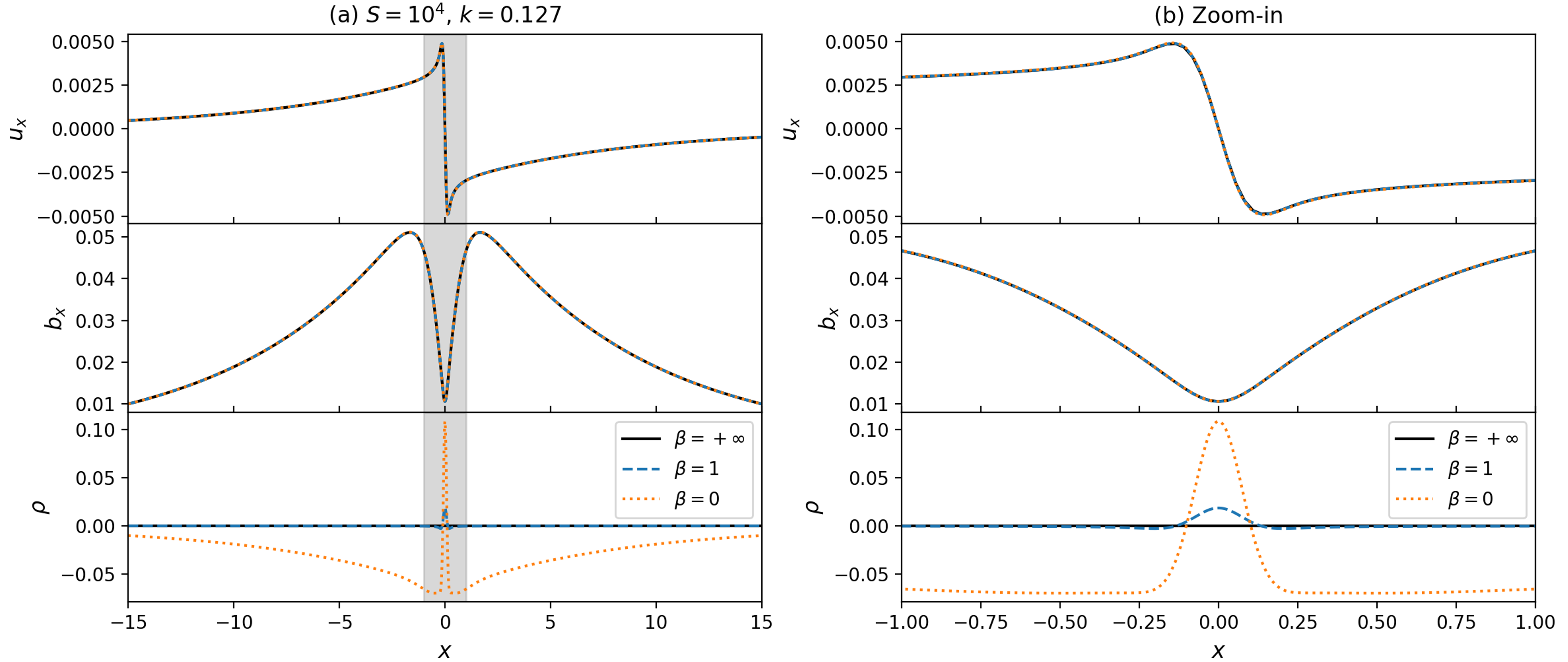}
    \caption{Solved eigenfunctions for $S=10^4$ and $k=0.127$ (the most unstable mode for $S=10^4$). The right plot is a zoom-in view of the region marked by gray shading in the left plot. In each plot, the panels from top to bottom show $u_x$, $b_x$ and $\rho$ respectively. Black lines correspond to the incompressible case ($\beta \rightarrow +\infty$), blue dashed lines to $\beta=1$, and orange dotted lines to $\beta=0$. }
    \label{fig:eigenfunctions}
\end{figure}

\subsection{Boundary condition}
To numerically solve Equation~(\ref{eq:full_ODE}) and determine the eigenvalue $\gamma$, we need to specify the boundary conditions of the system.
Far from the current sheet, the background magnetic field becomes nearly-uniform, with $B_z \approx Const$ and $B_y \approx 0$.
The perturbation amplitudes should therefore decay with increasing distance from the center of the current sheet.
Assuming uniform background fields and writing the perturbed fields in the form $f(x) = \Bar{f} e^{\alpha x}$,
Equation~(\ref{eq:full_ODE}) gives
\begin{equation}
    \left[ \begin{array}{cc}
        k B & \gamma - \frac{1}{S} \left( \alpha^2 - k^2 \right) \\
      \gamma - \frac{1 + k F}{\gamma} \beta \alpha^2  & \frac{B}{k} \left( \alpha^2 - k^2 \right)  
    \end{array} \right]  \left( \begin{array}{c}
         \Bar{u}_x \\
         \bar{b}_x 
    \end{array} \right) = 0
\end{equation}
where $B$ is the asymptotic value of $B_z$ and should be $\pm 1$, 
\begin{equation}
    F = - \frac{k\beta }{\gamma^2 + k^2 \beta}
\end{equation}
such that $u_z = F  u_x^\prime$ (Equation~(\ref{eq:full_ODE_uz})). 
Requiring the determinant of the matrix on the left-hand side to vanish gives the equation for $\alpha(k,\gamma|\beta)$:
\begin{equation}\label{eq:alpha}
\begin{aligned}
p_4 \alpha^4  + p_2 \alpha^2 +  p_0 = 0
\end{aligned}
\end{equation}
with 
\begin{equation}
    \begin{aligned}
        p_4 & = \frac{C \beta  } {S \gamma} \\
        p_2 & = - \left( B^2 + \frac{\gamma}{S} \right) - C \beta \left( 1 + \frac{k^2}{S \gamma} \right) \\
        p_0 & = \gamma^2 + k^2 \left(B^2 + \frac{\gamma}{S} \right)
    \end{aligned}
\end{equation} 
where
\begin{equation}
    C = 1 + kF = \frac{\gamma^2}{\gamma^2 + k^2 \beta }.
\end{equation}
In the classical incompressible case, $\alpha = \pm k$, with the sign of $\alpha$ chosen to ensure that the perturbations decay away from the current sheet.
Equation~(\ref{eq:alpha}) gives two solutions for $\alpha^2$, one of which is close to $k^2$; we use this solution to specify the boundary condition.

\section{Results}\label{sec:results}
We use the boundary-value-problem solver implemented in the SciPy package \citep{2020SciPy-NMeth} to solve Equation~(\ref{eq:full_ODE}). 
The solver employs a fourth-order collocation algorithm with a damped Newton method, and can simultaneously solve for unknown parameters ($\gamma$ in this case).
The computational domain is set to $x/a\in[-15,15]$.
The solver adaptively adjusts both the locations and the number of grid points, typically between 1000--4000 points, to achieve convergence for a prescribed tolerance. For a system $y^\prime = f(x,y)$, the tolerance is defined through the normalized residual criterion:
\begin{displaymath}
    \varepsilon   \geq \mathrm{RMS}\left( \frac{y^\prime - f(x,y)}{1 + |f(x,y)|} \right).
\end{displaymath}
We adopt $\varepsilon = 10^{-3}$ throughout this study. 
To assess the sensitivity of the results to this choice, we also performed calculations with $\varepsilon = 5 \times 10^{-4}$ and $10^{-2}$ for $S=10^6$ and found that the resulting growth rates were insensitive to the tolerance over this range.

\begin{figure}
    \centering
    \includegraphics[width=0.5\linewidth]{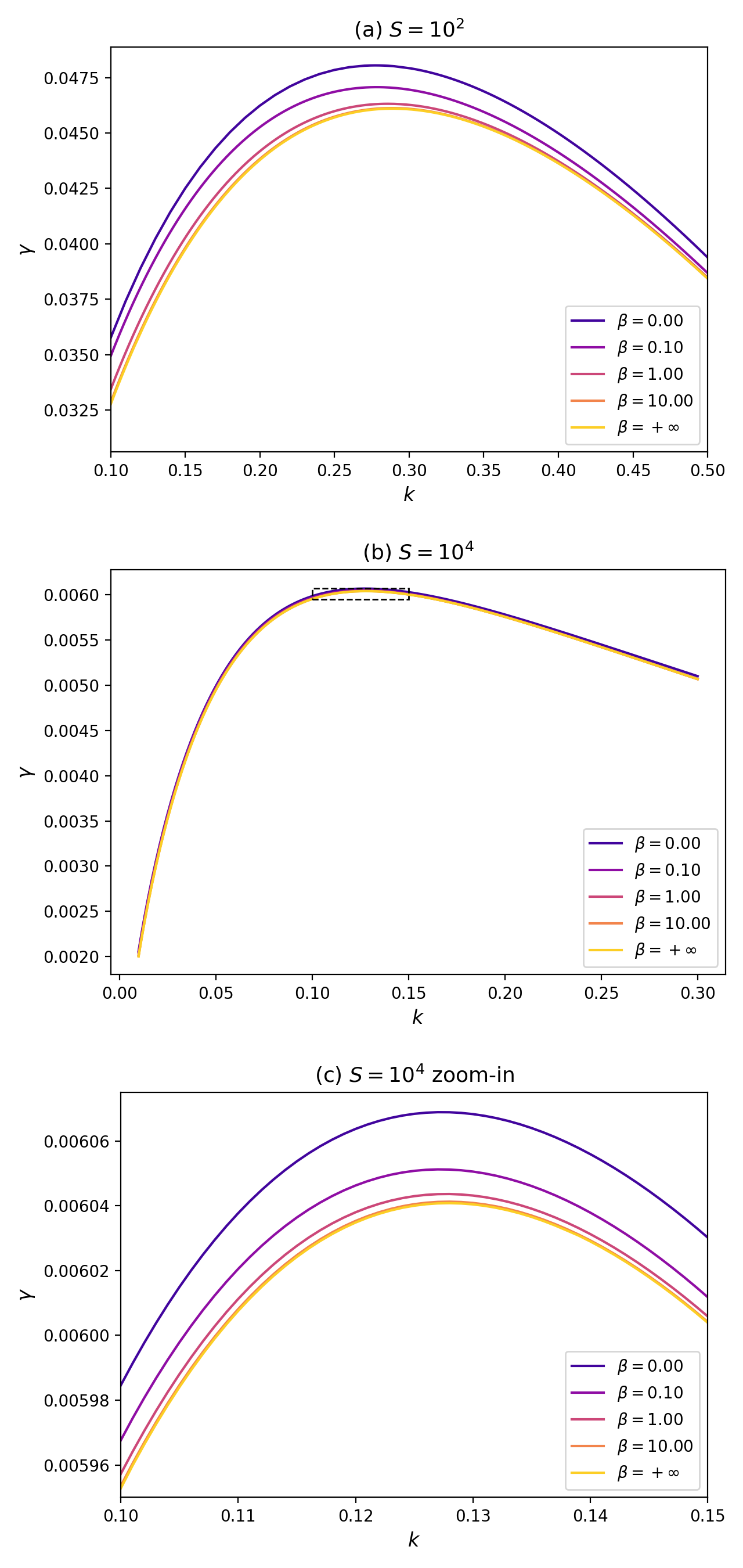}
    \caption{Dispersion relation $\gamma-k$ for different $\beta$. Panel~(a): $S=10^2$. Panel~(b): $S=10^4$. Panel~(c): Zoom-in of the region marked by the black box in panel~(b).}
    \label{fig:disp_diff_beta}
\end{figure}

Fig.~\ref{fig:eigenfunctions} shows an example of the solved eigenfunctions $u_x$, $b_x$, and $\rho$ for $S=10^4$ and $k=0.127$.
Column~(b) is a zoom-in view of column~(a), as indicated by the gray shading.
In each panel, we show three curves corresponding to three values of $\beta$: the incompressible limit ($\beta=+\infty$), $\beta=1$, and $\beta=0$.
The velocity and magnetic field perturbations are only weakly affected by variations in $\beta$.
In contrast, the density perturbation depends strongly on $\beta$. As $\beta$ decreases toward zero, a strong density perturbation develops within the inner layer.
From the ranges of the vertical axes, one can see that, in the limit $\beta \rightarrow 0$, the amplitude of $\rho$ becomes significantly larger than those of $u_x$ and $b_x$, which is associated with the sharp variation of $u_x$ across the inner layer.

\begin{figure}
    \centering
    \includegraphics[width=\linewidth]{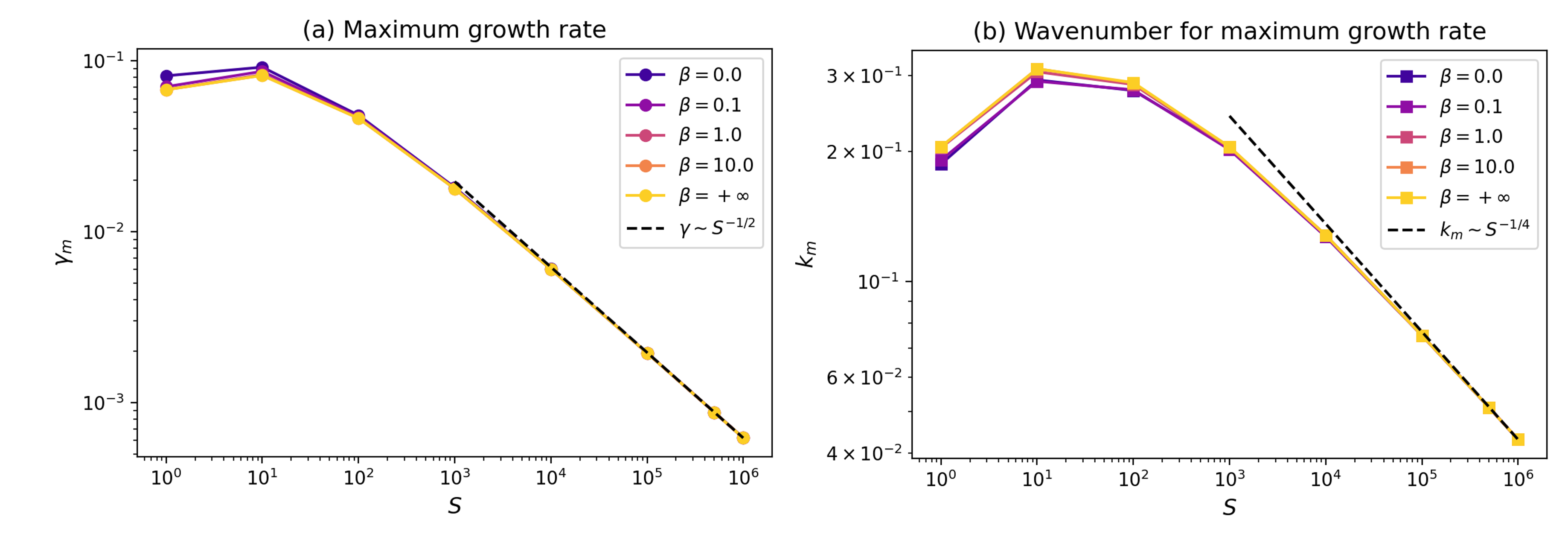}
    \caption{(a) Maximum growth rate as a function of $S$ for different $\beta$. (b) Wavenumber for maximum growth rate as a function of $S$ for different $\beta$.}
    \label{fig:max_gamma_km_vs_S}
\end{figure}

In Fig.~\ref{fig:disp_diff_beta}, we show the calculated dispersion relation $\gamma-k$ for $S=10^2$ (panel~(a)) and $S=10^4$ (panels~(b) \&~(c)). Here panel~(c) is a zoom-in view of the region enclosed by the black box in panel~(b).
In each panel, different curves correspond to different values of $\beta$: from dark to light, $\beta$ increases from zero to infinity.
In general, the growth rate increases as $\beta$ decreases, and the increase of $\gamma$ is more pronounced for $S=10^2$ than for $S=10^4$.
For each combination $(S,\beta)$, we calculate the maximum growth rate $\gamma_m$ and the corresponding wavenumber $k_m$.
In Fig.~\ref{fig:max_gamma_km_vs_S}, we plot $\gamma_m-S$ (panel~(a)) and $k_m-S$ (panel~(b)) for different values of $\beta$.
For reference, we plot $\gamma \propto S^{-1/2}$ and $k \propto S^{-1/4}$, which are the asymptotic scaling relations for the incompressible tearing mode, as black dashed lines.
Clearly, for large $S$ ($S \geq 10^4$), all curves converge, indicating that $\beta$ has only a negligible influence on the maximum tearing mode growth rate.
For smaller $S$, the curves diverge, and the increase in $\gamma_m$ associated with  decreasing $\beta$ becomes more significant.
$k_m$ also varies with $\beta$: smaller $\beta$ corresponds to smaller $k_m$, i.e. longer wavelength.

\begin{figure}
    \centering
    \includegraphics[width=0.5\linewidth]{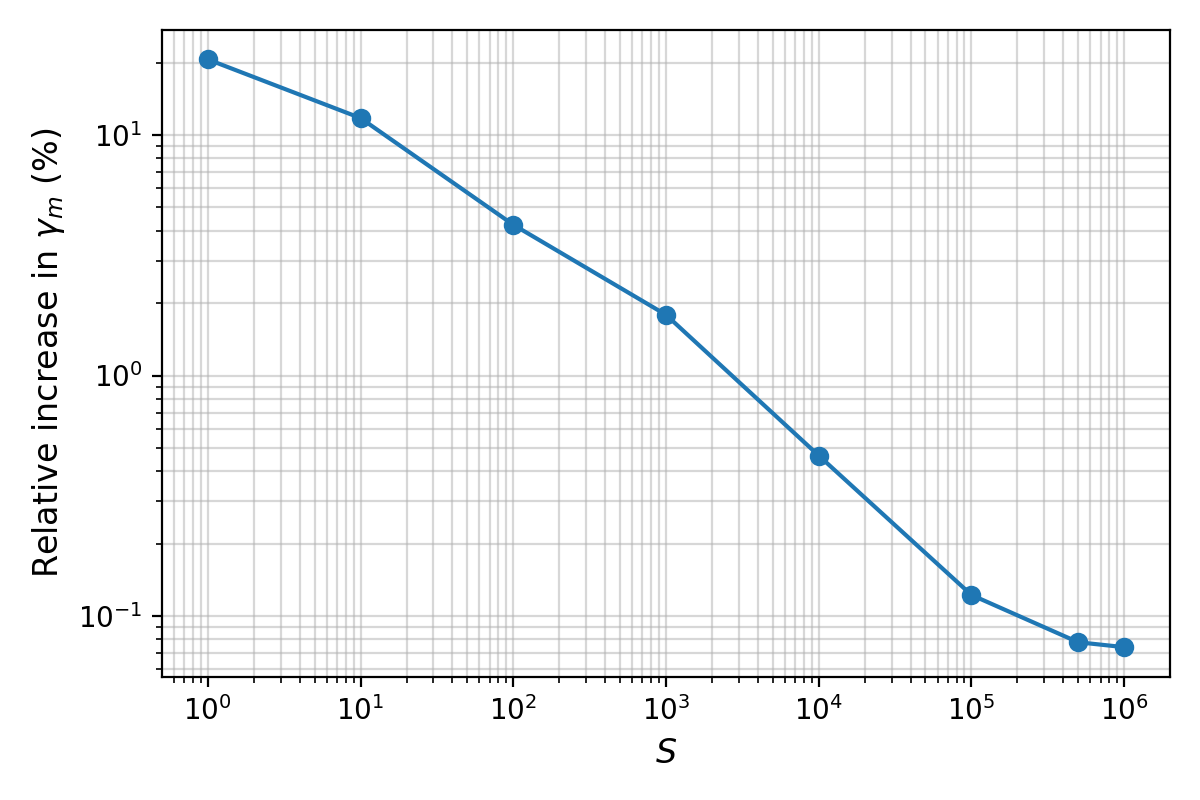}
    \caption{Relative increase of $\gamma_m$ as $\beta$ decreases from $+\infty$ to zero (Equation~(\ref{eq:delta_gamma})), as a function of $S$. }
    \label{fig:relative_increase_of_gamma_m}
\end{figure}

We quantify the relative increase in $\gamma_m$ as $\beta$ decreases from $+\infty$ to zero, i.e.
\begin{equation}\label{eq:delta_gamma}
    \Delta(S) = \frac{\gamma_m(S,\beta=0) - \gamma_m(S,\beta=+\infty)}{\gamma_m(S,\beta=+\infty)},
\end{equation}
and show the result in Fig.~\ref{fig:relative_increase_of_gamma_m}.
Consistent with Fig.~\ref{fig:max_gamma_km_vs_S}, the relative increase in $\gamma_m$ is negligible for $S>10^4$, remaining below 1\%, but becomes more significant at smaller $S$: for $S=10^2$, $\Delta$ is approximately 4\%, while for $S=10$, $\Delta$ can exceed 10\%.

\begin{figure}
    \centering
    \includegraphics[width=0.5\linewidth]{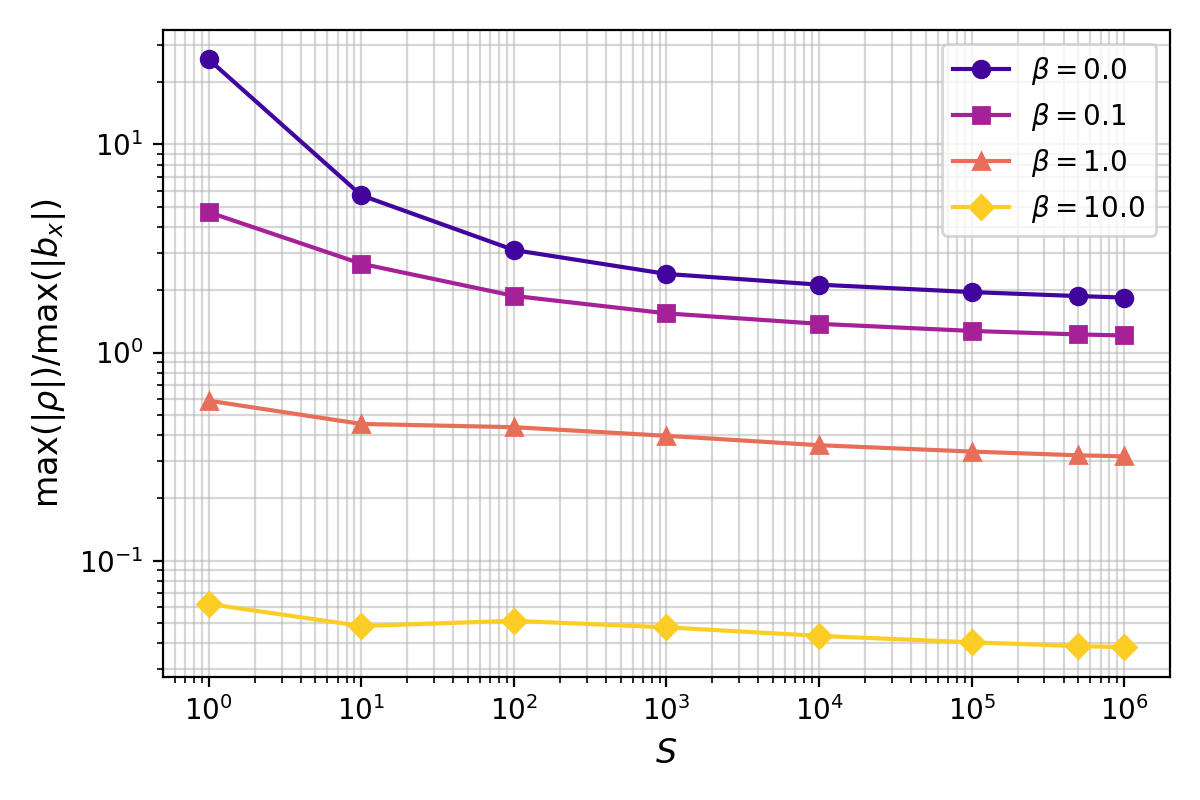}
    \caption{Maximum amplitude of density fluctuation divided by maximum amplitude of $b_x$ for the most unstable mode as a function of $S$. Different curves correspond to different $\beta$. }
    \label{fig:density_fluctuation_amplitude}
\end{figure}

As mentioned above, the density perturbation can be very strong in the low-$\beta$ regime.
In Fig.~\ref{fig:density_fluctuation_amplitude}, we show the ratio of the maximum amplitudes (i.e., the peaks of the eigenfunctions) of $\rho$ and $b_x$ for the most unstable mode as a function of $S$.
The density perturbation decreases with increasing $S$, although the decrease is relatively weak.
In contrast, the density perturbation is highly sensitive to $\beta$, with smaller $\beta$ producing stronger density perturbations. 
Therefore, in the compressible tearing mode, strong density perturbations can develop near the center of the current sheet. These density perturbations may influence the transition from the linear to the nonlinear stage of the instability.

\section{Conclusion}\label{sec:conclusion}
In this study, we derived the governing equations for the tearing mode from the fully-compressible resistive-MHD equations and computed its linear growth rate in a Harris-type force-free current sheet using a boundary-value-problem solver.
Our results show that compressibility generally increases the tearing mode growth rate.
This increase is weak at large Lundquist numbers ($S >10^4$) but becomes non-negligible at small Lundquist numbers ($S<10^2$).
Nevertheless, even though the tearing mode growth rate depends only weakly on plasma-$\beta$ at large $S$, the associated density perturbation can be strong and may exceed the magnetic field perturbation in relative amplitude as $\beta$ decreases.
Thus, in low-$\beta$ current sheets, such as those in the lower solar corona, the growth of tearing instability can generate strong density perturbations near the center of the current sheet.
These strong compressive fluctuations may influence the subsequent nonlinear evolution and thermodynamics of the current sheet.

The increase of tearing mode growth rate at small Lundquist numbers suggests, within the assumptions of the present linear model, that compressibility may lower the critical Lundquist number for the triggering of plasmoid instability relative to the incompressible case. 
This trend is opposite to that reported in previous compressible-MHD simulations \citep{ni2012effects,baty2014effect}, which found a lower critical Lundquist number at higher plasma-$\beta$.
We emphasize that these simulations do not employ the same equilibrium configuration as that considered here. In particular, thermal-pressure gradients contribute to maintaining force balance in those models, resulting in a nonuniform plasma-$\beta$ across the current sheet, whereas the present force-free equilibrium has uniform thermal pressure. This difference in equilibrium structure may contribute to the contrasting plasma-$\beta$ dependence.

Previous linear studies also do not provide a direct one-to-one comparison with the present model. 
\citet{tachi1985radiative} found that compressibility slightly reduces the tearing-mode growth rate, but their calculation employed a radiative-MHD model that includes radiative cooling, Joule heating, and temperature-dependent resistivity, rather than the adiabatic closure adopted here. 
\citet{de2024modification} found a plasma-$\beta$ dependence of the tearing-mode growth rate using the LEGOLAS eigenvalue solver, but their force-free equilibrium differs from ours and the $\beta$ dependence was examined only at a fixed wavenumber, rather than by identifying the most unstable mode over the full dispersion relation for each $S$ and $\beta$. 
These differences in equilibrium configuration, thermodynamic treatment, and numerical methodology make a direct quantitative comparison difficult.

We therefore do not regard the discrepancy with previous results as fully resolved. 
A definitive comparison would require calculations using similar equilibrium configuration and physical assumptions. 
In particular, determining the linear tearing-mode growth rate for equilibria with nonuniform thermal pressure, and conducting compressible-MHD simulations initialized with the same force-free equilibrium considered here, would help clarify the origin of the discrepancy. 
These investigations, however, are beyond the scope of the present study and are left for future work.

\section{Acknowledgment}
This work is supported by NSF SHINE \#2548299 and NASA ECIP \#80NSSC26K0321.

Competing interests: The author(s) declare none.

\section{Use of Artificial Intelligence (AI) Tools}
ChatGPT 5.6 Sol was used solely for language editing and polishing of the manuscript.


\bibliographystyle{jpp}

\bibliography{references}

\end{document}